\documentclass[aps,prb,reprint,twocolumn,superscriptaddress,citeautoscript,amsmath,amssymb,floatfix]{revtex4-2}
\usepackage{graphicx}
\usepackage{times}
\usepackage{bm}
\usepackage{color}
\usepackage{gensymb}
\usepackage{dcolumn}
\usepackage{xfrac}
\usepackage{booktabs}
\usepackage[colorlinks=true,citecolor=blue,linkcolor=blue,pdftex,hypertexnames=false,urlcolor=blue,linktocpage=true]{hyperref}
\usepackage{setspace}
\usepackage[none]{hyphenat}

\usepackage{adjustbox}
\newcolumntype{P}[1]{>{\centering\arraybackslash}p{#1}}
\usepackage{lipsum}

\usepackage{textcomp}

\newcommand{\SrNiCoP}[0]{$\text{Sr}(\text{Ni}_{1-x}\text{Co}_x)_{2}\text{P}_{2}$}
\newcommand{\SrCoNiP}[0]{$\text{Sr}(\text{Co}_{1-x}\text{Ni}_x)_{2}\text{P}_{2}$}
\newcommand{\SrNiCoAs}[0]{$\text{Sr}(\text{Ni}_{1-x}\text{Co}_x)_{2}\text{As}_{2}$}
\newcommand{\SrCoPGe}[0]{$\text{Sr}\text{Co}_{2}(\text{P}_{x}\text{Ge}_{1-x})_{2}$}

\newcommand{\SrNiP}[0]{$\text{Sr}\text{Ni}_{2}\text{P}_{2}$}

\newcommand{\SrCoP}[0]{$\text{Sr}\text{Co}_{2}\text{P}_{2}$}

\newcommand{\ThCrSi}[0]{$\text{Th}\text{Cr}_{2}\text{Si}_{2}$}

\newcolumntype{d}[1]{D{.}{.}{#1}}

\begin{document}
\title{Doping-induced Ferromagnetic order and its unusual evolution to Helical Antiferromagnetic Order in Sr(Ni$_{1-x}$Co$_x$)$_2$P$_2$}

\author{A. Sapkota}
\affiliation{Ames National Laboratory, U.S. DOE, Iowa State University, Ames, Iowa 50011, USA}
\affiliation{Department of Physics and Astronomy, Iowa State University, Ames, Iowa 50011, USA}

\author{J. Schmidt}
\affiliation{Ames National Laboratory, U.S. DOE, Iowa State University, Ames, Iowa 50011, USA}
\affiliation{Department of Physics and Astronomy, Iowa State University, Ames, Iowa 50011, USA}

\author{J.~M.~Wilde}
\affiliation{Ames National Laboratory, U.S. DOE, Iowa State University, Ames, Iowa 50011, USA}
\affiliation{Department of Physics and Astronomy, Iowa State University, Ames, Iowa 50011, USA}

\author{L.-L. Wang}
\affiliation{Ames National Laboratory, U.S. DOE, Iowa State University, Ames, Iowa 50011, USA}
\affiliation{Department of Physics and Astronomy, Iowa State University, Ames, Iowa 50011, USA}

\author{W. Tian}
\author{M. Matsuda}
\affiliation{Neutron Scattering Division, Oak Ridge National Laboratory, Oak Ridge, Tennessee 37831, USA}

\author{A. Kreyssig}
\affiliation{Ames National Laboratory, U.S. DOE, Iowa State University, Ames, Iowa 50011, USA}
\affiliation{Department of Physics and Astronomy, Iowa State University, Ames, Iowa 50011, USA}
\affiliation{Institute for Experimental Physics 4, Ruhr-Universit\"{a}t Bochum, 44801 Bochum, Germany}

\author{S.~L.~Bud'ko}
\affiliation{Ames National Laboratory, U.S. DOE, Iowa State University, Ames, Iowa 50011, USA}
\affiliation{Department of Physics and Astronomy, Iowa State University, Ames, Iowa 50011, USA}

\author{P.~C.~Canfield}
\affiliation{Ames National Laboratory, U.S. DOE, Iowa State University, Ames, Iowa 50011, USA}
\affiliation{Department of Physics and Astronomy, Iowa State University, Ames, Iowa 50011, USA}

\date{\today}

\begin{abstract}

\SrNiP{} represents a unique case of collapsed structural phase (one-third collapsed; where one out of every three P--P pairs forms a bond) in the $A(TM)_2X_2$ series of compounds ($A$ = an alkali metal, alkaline earth, or rare earth; $TM$ = a transition metal; and $X$ = Pnictogen). Furthermore, Co-doping studies aimed at understanding the interrelationship between this unusual bonding motif and the resulting physical properties produced a magnetically rich phase diagram, specifically on the cobalt rich side of the phase diagram. However, important questions remained regarding the detailed nature of the magnetic ground states. To address these issues, we performed single-crystal neutron diffraction on \SrNiCoP{} with compositions \(x = 0.88\), \(0.94\), and \(0.97\). For \(x = 0.88\) and \(0.94\), the measurements reveal incommensurate helical magnetic order with a doping-dependent propagation vector \((0,0,\tau)\), similar to that observed in \SrNiCoAs{}. In contrast, the \(x = 0.97\) composition shows clear signatures of a ferromagnetically ordered ground state, resolving the earlier ambiguity regarding the nature of the low-temperature phase. Furthermore, our results highlight the subtle balance between these competing ground states, whose evolution does not appear to be fully captured by the conventional frameworks of either itinerant or local-moment Heisenberg models typically applied to related 122 systems.

\end{abstract}

\maketitle

\section{Introduction}
The $A(TM)_2X_2$  ($A$ = an alkali metal, alkaline earth, or rare earth; $TM$ = a transition metal; and $X$ = Pnictogen; commonly referred as "122") series of compounds,  with the \ThCrSi{} structure, have proven to be interesting due to the rich phase diagrams with tunable and diverse structural, magnetic, and electronic phases \cite{Canfield_2009,Ni_2010,Gati_2012,Li_2019,Gati_2020,Trovarelli_2000}. The intertwined structural, magnetic, and electronic ground states are highly sensitive to external tuning parameters such as pressure and chemical substitution. One of the most intriguing consequences of such tuning is the breaking or formation of interlayer $X$–$X$ bonds \cite{Hoffman_1985,Juan_2023, Juan_2025_Rh}, which in turn strongly influences the mechanical \cite{Syperk_2017,Xiao_2021}, electronic \cite{Gati_2012}, and magnetic \cite{Kreyssig_2008,Jia_2009,Jia_2011} properties of these materials. 

Based on bonding or antibonding character of $X-X$\cite{Hoffman_1985}, also quantified in terms of $X-X$ distances, their tetragonal $I4/mmm$ chemical structure has been typically classified as collapsed tetragonal (cT; shorter $X-X$ distance with bonding) and uncollapsed tetragonal (ucT; longer distance with antibonding) phases \cite{Kreyssig_2008}. Further classification such as half-collapsed tetragonal phase has been introduced in the related CaKFe$_4$As$_4$ family of materials, in which As atoms bond across every other Ca-layer\cite{Kaluarachchi_2017,Borisov_2018}.

\begin{figure}[htbp]
    \centering
    \includegraphics[width=\columnwidth]{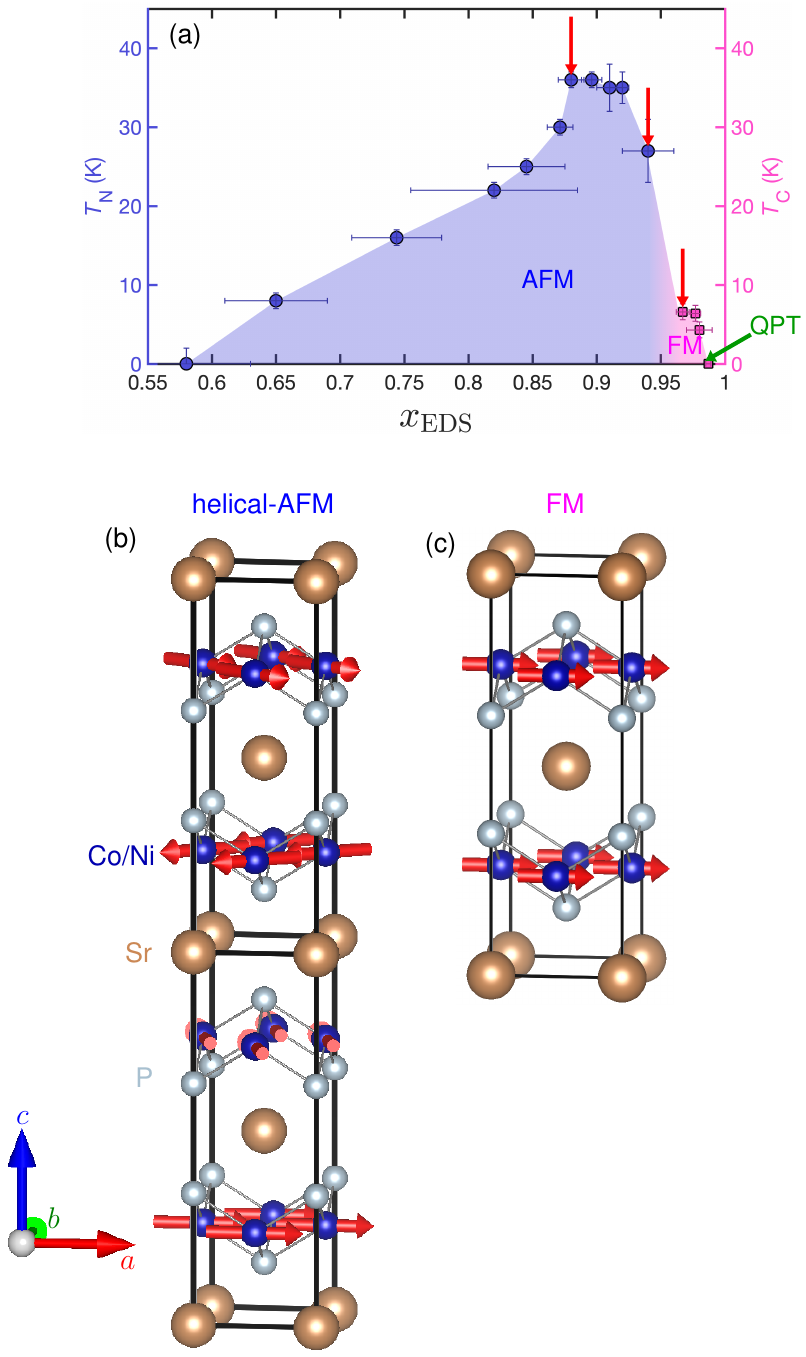}
    \caption{Magnetic phase diagram and magnetic structures corresponding to \SrNiCoP{}. (a) The magnetic phase diagram of \SrNiCoP{} with N\'eel temperatures ($T_\text{N}$) in blue for the antiferromagnetic phase (AFM), and Curie temperatures ($T_\text{C}$) in magenta for the ferromagnetic phase (FM). Critical temperatures determined from bulk magnetic susceptibility in Ref.~\cite{Juan_2023} are shown as filled markers. Neutron diffraction samples are shown with red arrows and quantum phase transition (QPT) near \SrCoP\ end member with green arrow. (b) Helical antiferromagnetic structure with the propagation vector, $(0\,0\,\tau)$, for $\tau \sim 0.6$ r.l.u., which shows ferromagnetically aligned layers with the ordered magnetic moment perpendicular to $\bm{c}$. Ferromagnetic planes are stacked along $\bm{c}$ with a turn angle of $\approx107\degree$ with respect to the nearest layers. (c) Ferromagnetic structure with ordered magnetic moment in the ab-plane. In (c), the direction of $\bm{\mu}$ in the plane was arbitrarily chosen to be $\parallel [1\,0\,0]$. The magnetic structures were created using VESTA \cite{Vesta_2011}.
    }
    \label{Phase-Diagram}
\end{figure}

In this regard, \SrNiP{} represents a unique case: it undergoes a one-third collapsed orthorhombic (tcO) transition below 325 K\cite{Juan_2023}, adopting the $Immm$ symmetry. Remarkably, only one out of every three P–P rows forms bonds across the Sr layers, shown in Fig. 1 of Ref.~\citenum{Keimes_1997}$-$a phenomenon whose microscopic origin remains unexplained. To understand the interrelationship of this unique bonding with its physical properties, such as magnetism and superconductivity, various systematic chemical substitutional studies were carried out in Ref.~\citenum{Juan_2023} and ~\citenum{Juan_2025_Rh}. In particular, the series \SrNiCoP{} revealed that Co-substitution tunes the ucT ($I4/mmm$) to tcO ($Immm$) transition of parent \SrNiP{}. The Co-substitution rapidly suppresses the transition by $x \approx 0.1$ leaving the system in ucT crystal structure, down to base temperature, beyond $x > 0.1$. Additionally, magnetic ordering emerges around $x \geq 0.65$ and vanishes again near $x \leq 0.99$, consistent with the paramagnetic \SrCoP{} end member. Furthermore, the magnetic phases have been identified as antiferromagnetic (AFM) and ferromagnetic (FM) as shown in Fig.~\ref{Phase-Diagram}(a) for the Co rich side of the phase diagram. However, as discussed in Ref.~\citenum{Juan_2023}, ambiguity remains regarding the true nature of both magnetic phases$-$whether the AFM state is spin-glassy and the FM phase is bulk$-$thereby necessitating neutron diffraction measurements to determine their magnetic ground states. Another aspect of this series is the deviation from FM ground state with only a small change in doping $x$, occurring without any significant structural modification. 

To elucidate the microscopic nature of the magnetic ordering and to capture the evolution of the ground states, we performed single-crystal neutron diffraction measurements on compositions $x = 0.88$ (AFM phase), $0.94$ (AFM–FM boundary), and $0.97$ (FM phase). The diffraction results for $x = 0.88$ and $0.94$ reveal helical magnetic ordering with a doping-dependent incommensurate propagation vector of $(0,0,\tau)$, similar to that observed in \SrNiCoAs{}\cite{Wilde_2019}. For the composition $x = 0.97$, the neutron diffraction results confirm the presence of FM ground state, clarifying the previous ambiguity. Furthermore, the observed change in magnetic ground states were discussed in the light of both itinerant model and competing exchange interactions in the Heisenberg models.

\section{Experimental Methods}

Neutron diffraction measurements were performed on single-crystals of \SrNiCoP{} for $x=0.88(1)$, $x=0.94(2)$, and $x=0.97(5)$ as shown in Fig.~\ref{Phase-Diagram}(a). The crystals were grown using the Sn-flux method, as detailed in Ref~\cite{Juan_2023}, and the composition was determined by energy dispersive X-ray spectroscopy (EDS). Additionally, Cu-doped samples, discussed below in the Appendix~\ref{CuComp}, were also grown using the Sn-flux. A summary of both crystal growth details is given below in the Appendix \ref{SCGC}. The mass of the single-crystals used for the neutron measurements was 7.6(1), 6.6(1), and 7.7(1)~mg for $x=0.88$, 0.94, and 0.97, respectively. The $x=0.88$ crystal was measured at the VERITAS triple-axis spectrometer at the High Flux Isotope Reactor, Oak Ridge National Laboratory. The $x=0.94$ and 0.97 single-crystals were measured at the neighboring PTAX spectrometer at the High Flux Isotope Reactor, Oak Ridge National Laboratory. For each case samples were sealed in an aluminum can containing helium as an exchange gas, which was then attached to the cold head of a closed-cycle helium refrigerator. Scattering data for for all studies are described using the reciprocal lattice units $H$, $K$, and $L$ for the tetragonal unit cell with the \ThCrSi{} type structure, and lattice parameters of $a=b\approx{}3.8$~\AA{}, and $c\approx{}11.6$~\AA{}. All samples were aligned such that the $(H\,H\,L)$ plane was coincident with the horizontal scattering plane of the instrument.

VERITAS was operated at a fixed incident energy of 14.46 meV using two pyrolytic graphite (PG) monochromators. Horizontal beam collimations (angular divergence of the beam)  of $40^\prime$--$40^\prime$--$40^\prime$--$80^\prime$ were used, with collimators located before the monochromator, between the monochromator and the sample, between the sample and the analyzer, and between the analyzer and the detector, respectively. Measurements on PTAX were operated at a fixed incident energy of 13.5 meV using two PG monochromators. The beam collimators for P-TAX are placed before the monochromator, between the sample and monochromator, between the sample and analyzer, between the analyzer and detector are $48^{\prime}-80^{\prime}-60^{\prime}-240^{\prime}$, respectively. Higher harmonics within the incident beam are significantly reduced by placing a PG filter before and after the second monochromator for both VERITAS and PTAX.

DC magnetization measurements were carried out on a Quantum Design Magnetic Property Measurement System (MPMS classic) superconducting quantum interference device (SQUID) magnetometer (operated in the range $1.8\ \text{K}\leq T \leq 300\ \text{K}$, $|H|\leq 50\ \text{kOe}$). Each sample was measured with the field applied perpendicular to the tetragonal $c$ axis.

Density functional theory (DFT) calculations were performed using the Perdew--Burke--Ernzerhof (PBE) exchange--correlation functional~\cite{Hohenberg1964,Kohn1965,PBE1996} within a plane-wave basis and the projector augmented-wave (PAW) method~\cite{Blochl1994}, as implemented in the Vienna \textit{Ab initio} Simulation Package (VASP)~\cite{Kresse1996CMS,Kresse1996PRB}. A kinetic energy cutoff of 400~eV and Gaussian smearing of 0.05~eV were employed. $\Gamma$-centered Monkhorst--Pack~\cite{Monkhorst1976} $k$-point meshes of $4\times4\times4$ and $6\times6\times4$ were used for the $3\times3\times1$ and $2\times2\times1$ supercells of SrCo$_2$P$_2$ with Ni alloying, respectively. These supercells correspond to alloy compositions of $x_{\mathrm{Co}} = 0.97$ and $0.94$, obtained by substituting one Co atom with Ni.

Previously in \cite{Shuyan_2025}, to describe the pressure-induced structural and magnetic phase stability of pristine SrCo$_2$P$_2$ using DFT calculations,  demonstrated the importance of reproducing the experimental $c/a$ ratio by applying a shift to the calculated pressure, referred to as the pressure-shift protocol. Here, we apply the same protocol to investigate the structural and magnetic phase stability of Sr(Co,Ni)$_2$P$_2$, where the evolution of the $c/a$ ratio is instead driven by Ni alloying. The effects of Ni alloying on the structural $c/a$ ratio and the stability of the ferromagnetic (FM) state were then evaluated under the same pressure-shift conditions. For accurate density-of-states (DOS) calculations, denser $k$-meshes were used together with the tetrahedron method~\cite{Blochl1994Tetra}, with the $k$-point density doubled relative to that used for total-energy calculations.

\section{Results}

\begin{figure}[htbp]
    \centering
    \includegraphics[width=\columnwidth]{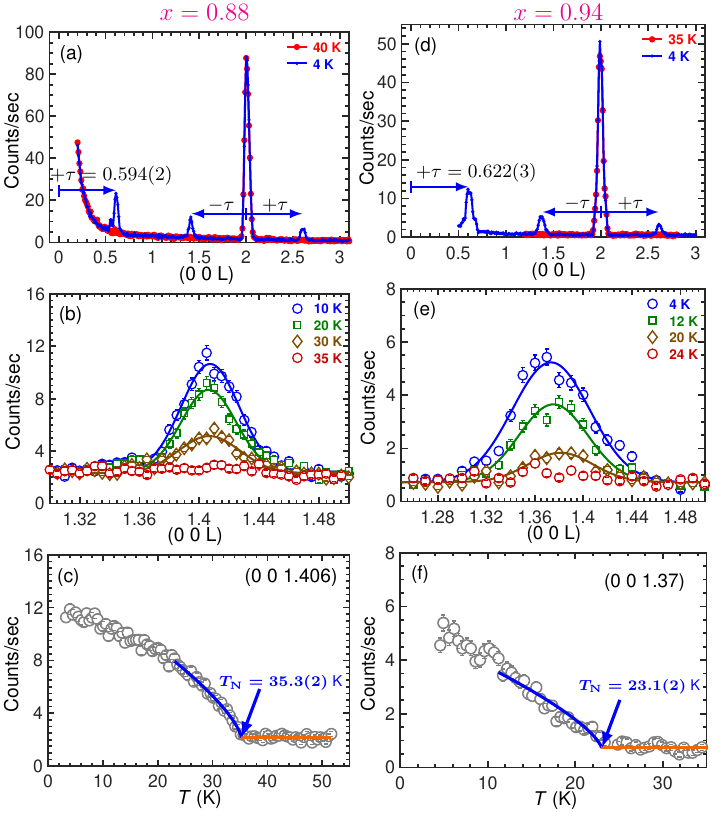}
    \caption{Single-crystal neutron diffraction data in the AFM region for $x=0.88$ and 0.94 of \SrNiCoP{}. (a), (d) $(0\,0\,L)$ scans showing additional Bragg peaks at low temperatures for $x=0.88$ and 0.94, respectively. (b), (e) $(0\,0\,L)$ scans of the incommensurate magnetic Bragg peak $(0\,0\,2)-\bm{\tau}$ at several temperatures for $x=0.88$ and 0.94, respectively. Solid lines are fits using Gaussian lineshapes. (c), (f) Intensity of the $(0\,0\,2)-\bm{\tau}$ Bragg peak (proportional to the ordered moment squared) vs temperature for $x=0.88$ and 0.94, respectively. Solid blue lines are from a power-law fit.}
    \label{AFM_Order_params}
\end{figure}

Figures~\ref{AFM_Order_params}(a)-(c) and \ref{AFM_Order_params}(d)-(f) show single-crystal neutron diffraction data consistent with antiferromagnetic order for $x=0.88$ and $0.94$, respectively. In Fig.~\ref{AFM_Order_params}(a) and \ref{AFM_Order_params}(d), satellite peaks appear at positions $(0\,0\,2)\pm\bm{\tau}$, and $(0\,0\,\tau)$ below the N\'eel temperature ($T_\text{N}$). The magnetic Bragg peaks appear at positions incommensurate to the chemical lattice with propagation vector $\bm{\tau}=(0\,0\,\tau)$, with $\tau=0.594(2)$, and $0.622(3)$ for $x=0.88$ and $0.94$, respectively. The most intense peaks were found at positions $(0\,0\,L)\pm\bm{\tau}$, where $L$ is even. Therefore, a significant component of the ordered magnetic moment must be in the plane, since the magnetic Bragg peaks at the position $(0\,0\,L)\pm\bm{\tau}$ are only sensitive to the component of $\bm{\mu}\perp\bm{c}$. The result is consistent with the magnetization data in Ref.~\citenum{Juan_2023}. In addition, the AFM phase appears long-range, since the width of the satellite peaks is similar to the width of the nuclear Bragg peak, as seen by comparison of the $(0\,0\,2)$ Bragg peak (FWHM = $0.055$ rlu) and the satellite peaks ($0.045-0.055$ rlu) in Fig.~\ref{AFM_Order_params}(a) and \ref{AFM_Order_params}(d), thereby ruling out any cluster glass-like phases. Figures \ref{AFM_Order_params}(b) and \ref{AFM_Order_params}(e) show that the propagation vector does not appear to be significantly temperature dependent, with shift being $\leq 0.008$ r.l.u., from $(0\,0\,L)$-scans of the $(0\,0\,2)-\bm{\tau}$ magnetic Bragg peak at several temperatures for $x=0.88$ and $0.94$, respectively. The value of $\tau$ appears to be far more dependent on $x$ from comparison of Fig.~\ref{AFM_Order_params}(b) and \ref{AFM_Order_params}(e), and as shown explicitly in Fig.~\ref{AFM_t-x_dep}(a) below. 


The temperature dependence of the intensity of the $(0\,0\,2)-\bm{\tau}$ magnetic Bragg peak (proportional to $|\bm{\mu}|^2$) is consistent with a second-order transition as shown in Fig.~\ref{AFM_Order_params}(c) and (f), for $x=0.88$ and $0.94$, respectively. The N\'eel temperature of the AFM phase for $x=0.88$ has the value $T_\text{N}=35.3(2)$~K, which was determined from a power-law fit as shown in Fig.~\ref{AFM_Order_params}(c). 
Similarly, from the power-law fit of the temperature dependence of $x=0.94$, the N\'eel temperature was determined to be $T_\text{N}=23.1(2)$~K as shown in Fig.~\ref{AFM_Order_params}(f). As seen in Fig.~\ref{Phase-Diagram}(a) the transition temperatures $T_\text{N}$ determined for $x=0.88$ and $x=0.94$ are consistent with critical temperatures determined from bulk magnetic susceptibility in our previous study \cite{Juan_2023}. In summary, the neutron diffraction data for $x=0.88$ and $0.94$ show the long-range incommensurate AFM order, with characteristics similar to the one observed in the related compound \SrNiCoAs{} \cite{Wilde_2019}.

The description of the AFM phase above is consistent with two types of antiferromagnetic structures: a collinear spin-density-wave (SDW) and a noncollinear helical magnetic structure. For a simple SDW the magnitude of $\bm{\mu}$ varies sinusoidally along $\bm{c}$, but the direction of $\bm{\mu}$ remains fixed. Whereas, for a noncollinear helix the magnitude of $\bm{\mu}$ remains constant, but the direction of $\bm{\mu}$ is allowed to vary within the ab plane. The SDW scenario would break the underlying tetragonal symmetry, potentially producing observable lattice distortions due to domain formation. Such distortions were not detected in our neutron diffraction data and, if present, are likely below the instrumental resolution. However, our neutron results with the recent anisotropic magnetic susceptibility measurements for the AFM phase, which show a metamagnetic transition and nonzero magnetic susceptibility as the temperature approaches zero for $\bm{H}\perp\bm{c}$ \cite{Juan_2023}, favors the helical magnetic structure. As shown in Fig.~\ref{Phase-Diagram} (b), the helical magnetic ordering is described as ferromagnetically aligned square layers that are antiferromagnetically arranged along $\bm{c}$.

\begin{figure}[htbp]
    \centering
    \includegraphics[width=\columnwidth]{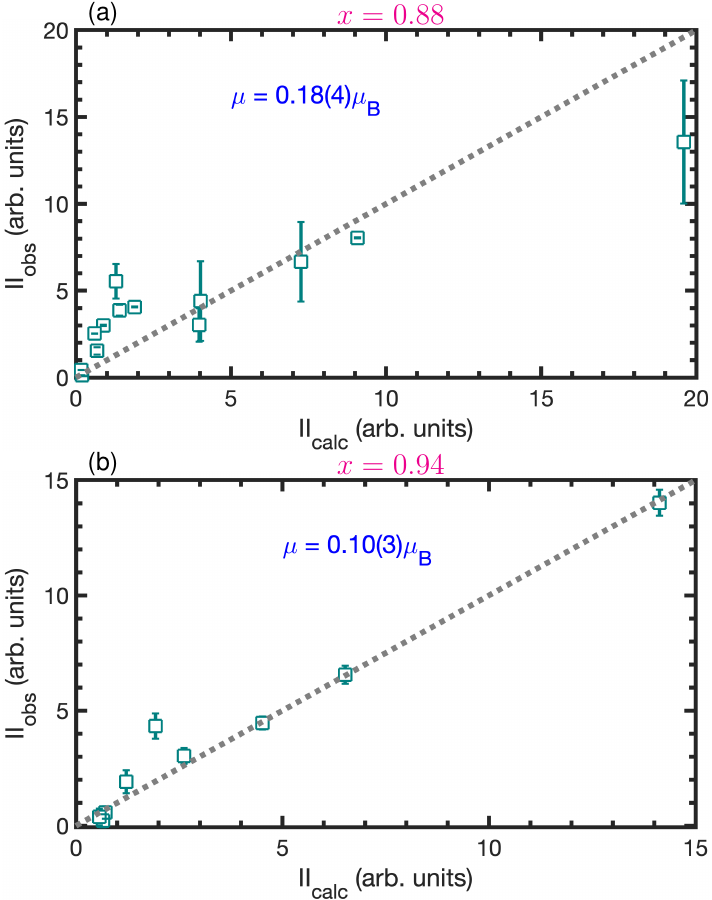}
     \caption{Comparison between the measured integrated intensities ($II_{\mathrm{obs}}$) and calculated integrated intensities ($II_{\mathrm{calc}}$) of the incommensurate magnetic reflections for  $x = 0.88$ (a) and $x = 0.94$ (b).  The calculated intensities were obtained using \textsc{FullProf} based on the helical magnetic structure shown in Fig.~\ref{Phase-Diagram}(b). The dashed line represents the best linear fit, whose slope corresponds to the squared ordered magnetic moment, yielding an ordered moment of $\mu = 0.18(3)$ and $0.10(3)\,\mu_{\mathrm{B}}$, respectively.}
    \label{MMD_Fullprof}
\end{figure}

Additionally, the propagation vector of the helix corresponds to a turn angle $\phi$ between the neighboring planes of magnetic moments. Hence, from it, we determined the $\phi$ to be $106.9(4)^\circ$ and $112.0(4)^\circ$ for $x = 0.88$ and $0.94$, respectively. Furthermore, we estimated the ordered moment associated with the helical-AFM structure by comparing the experimental integrated intensities to simulated Bragg peak intensities generated using FULLPROF \cite{rodriguez1993fullprof} for the magnetic structure shown in Fig.~\ref{Phase-Diagram}(b). Figure~\ref{MMD_Fullprof} illustrates the agreement between the experimental and calculated magnetic peaks intensities for $x = 0.94$ with the proposed helical order. Moreover, the resulting ordered moments are estimated to be $\mu = 0.18(3)$ and $0.10(3)~\mu_\text{B}/$TM (where TM = Co+Ni in this case) for $x = 0.88$ and $0.94$, respectively. These values agree well with the corresponding saturated moments, $\mu_\text{s} = 0.150(1)$ and $0.092(1)~\mu_\text{B}/(\text{TM})$, reported in Ref.~\cite{Juan_2023}.

\begin{figure}[htbp]
    \centering
    \includegraphics[width=\columnwidth]{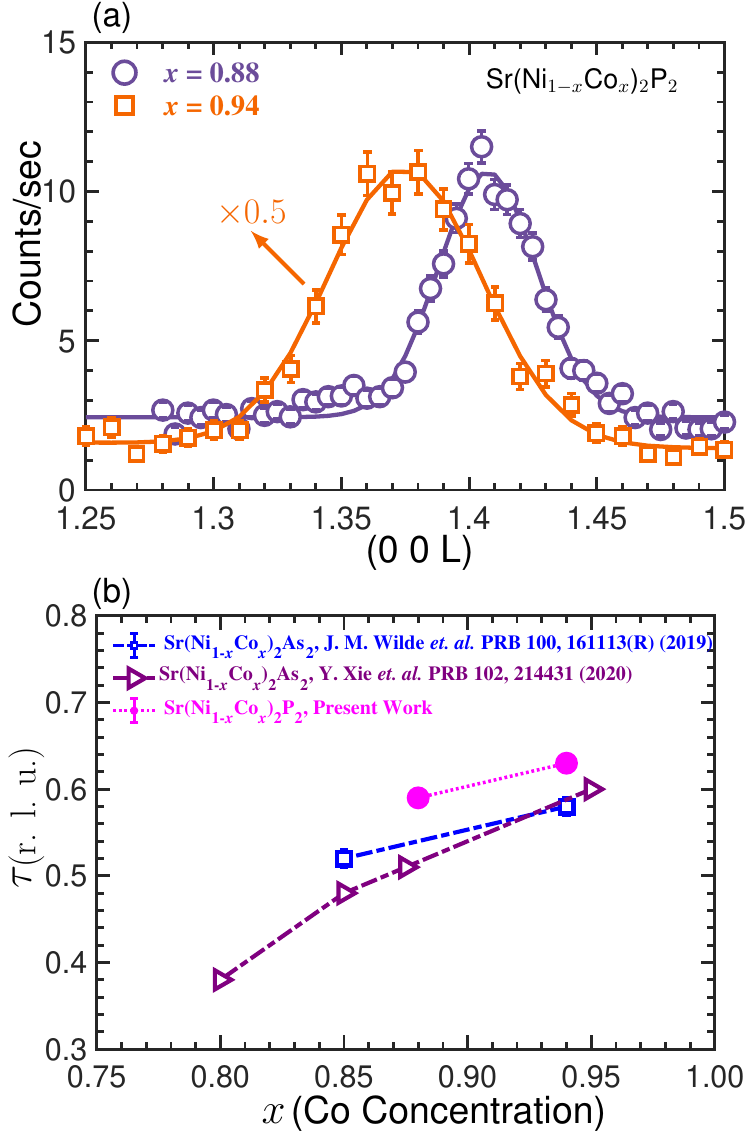}
    \caption{Evolution of the incommensurate antiferromagnetic propagation vector, $\bm{\tau}=(0\,0\,\tau)$, at the magnetic Bragg peak $(0\,0\,2)-\bm{\tau}$ with $x$. (a) $(0\,0\,L)$ scans showing incommensurate magnetic Bragg peaks for both $x=0.88$ and 0.94 single crystals. The decrease in the center of the peak in $L$ corresponds with an increase in $\tau$. (b) Plot showing the dependence of $\tau$ on the Co concentration for \SrNiCoP{} is consistent with results for \SrNiCoAs{} \cite{Wilde_2019,Xie_2020_SrNiCoAs,Li_2019_SrCoNiAs}.}
    \label{AFM_t-x_dep}
\end{figure}

As previously noted, the value of $\tau$ increases with increasing $x$ within the AFM phase of \SrNiCoP{}, as shown explicitly in Fig.~\ref{AFM_t-x_dep}(a) for the $(0\,0\,2)-\bm{\tau}$ magnetic Bragg peak. This behavior is similar to the helical antiferromagnetic order of \SrNiCoAs{} determined from several studies \cite{Wilde_2019,Xie_2020_SrNiCoAs,Li_2019_SrCoNiAs}. The AFM phase for \SrNiCoAs{} and \SrNiCoP{} appear to manisfest the same helix structure consisting of FM aligned Co square planes as shown in Fig.~\ref{Phase-Diagram}(b). 


\begin{figure}[htbp]
    \centering
    \includegraphics[width=\columnwidth]{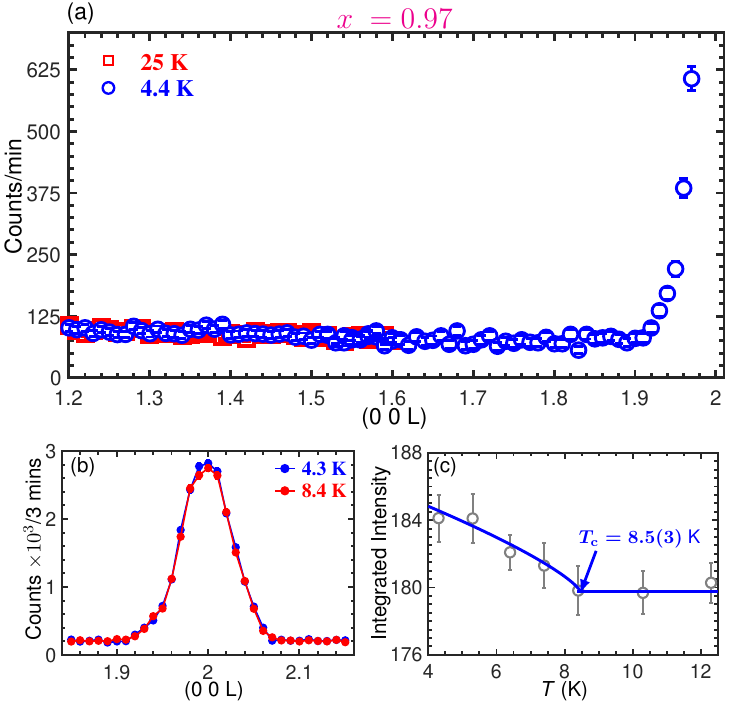}
    \caption{Neutron diffraction data within the FM phase for a single crystal of \SrNiCoP{} for $x=0.97$. (a) $(0\,0\,L)$ scans over a wide range. (b) $(0\,0\,L)$ scans of the $(0\,0\,2)$ Bragg peak showing an enhancement of intensity at low temperatures (c) Temperature dependence of the integrated intensity of the $(0\,0\,2)$ Bragg peak. The critical temperature was obtained from a power law fit of the data below 8.5~K where $\beta=0.38(3)$.}
    \label{FM_Order_params}
\end{figure}

Next, from our neutron diffraction results of $x = 0.97$, we confirm that long-range ferromagnetic order develops at high Co concentrations within the FM region of Fig.~\ref{Phase-Diagram}(a). As illustrated in Fig.~\ref{FM_Order_params}(a), wide $(0,0,L)$ scans collected above and below the critical temperature reveal no signatures of antiferromagnetic (AFM) order for this composition. Instead, a weak enhancement of the $(0,0,2)$ Bragg peak intensity, just beyond the error bars, is observed upon cooling below $T_\text{C}$, as shown in Fig.~\ref{FM_Order_params}(b). This increase in intensity is consistent with the onset of long-range ferromagnetic order with a small ordered moment, since below $T_\text{C}$ additional magnetic scattering is expected to superimpose on existing nuclear Bragg reflections. Furthermore, the enhancement was only observed for (0, 0, 2) peak, indicating the in-plane moment ferromagnet. Thus, when combined with the magnetization data reported in Ref.~\citenum{Juan_2023}, our neutron diffraction results confirm that the ferromagnetic order in this system is intrinsic.

To quantify this contribution, the $(0,0,2)$ Bragg peaks measured at temperatures above and below $T_\text{C}$, were fit with Gaussian lineshape to obtained the integrated intensities. Thus obtained integrated intensities, shown in Fig.~\ref{FM_Order_params}(c), exhibit a clear enhancement below $T_\text{C} = 8.5(3)$~K, consistent with a continuous (second-order) transition to a ferromagnetically ordered state. The $\sim2.4\%$ increase in intensity in Fig.~\ref{FM_Order_params}(c) is consistent with the enhancement expected from \textsc{FullProf} simulations using the ordered moment $\mu = 0.05~\mu_\text{B}/$TM reported for $x=0.97$ in Ref.~\cite{Juan_2023} and the magnetic structure shown in Fig.~\ref{Phase-Diagram}(c).

\section{Discussion \label{Discussion}}
Overall, an intriguing aspect of the magnetic phase diagram of the \SrNiCoAs{} series is the systematic evolution of magnetism from Stoner-enhanced paramagnetism (SEP) to ferromagnetism for small Ni-substitution and, further to a non-collinear helical state driven by only small changes in $x$. The thermodynamic and NMR\cite{Furukawa_2024} measurements clearly indicate the FM QPT at boundary of SEP-FM phase near the Co-end member. This behavior underscores the subtle balance of competing energy scales in this itinerant system, making the microscopic origin of the magnetism particularly intriguing and also challenging to unravel. As will be discussed below, standard theoretical frameworks, including those based on structural coupling ($X-X$ bonding)\cite{Jia_2009}, simple itinerant Stoner physics\cite{Stoner_1939}, or local-moment interactions\cite{Li_2019}, applied to other related pnictides, are either insufficient or incompatible with the present observations. This indicates that a more nuanced theoretical approach is required to fully capture the microscopic origin of magnetism in this series and suggests for the revision of the description of itinerant magnetism in the previously studied 122-materials.

Unlike many 122 Fe and Co pnictides \cite{Kreyssig_2008,Jia_2011,Li_2019}, the magnetic ordering in this series occurs without any accompanying significant structural changes and does not coincide with the uncollapsed-to-collapsed structural transition. Consequently, the mechanism proposed for \SrCoPGe{} \cite{Jia_2009}, which links ferromagnetism to a structural change involving the population of a $\sigma^{*}$ band (hybridized with Co 3d bands) and the breaking of $P-P$ bonds, does not seem applicable here. Given that the itinerant nature of the magnetism in this system is well established, as the magnetic ordering and moment formation are well described by Takahashi’s spin fluctuation theory \cite{Takahashi_1986}, considering the Stoner criterion, $I \times N(E_\text{F}) > 1$, for magnetic instability is a natural starting point. Here, $I$ is the Stoner interaction parameter and $N(E_\text{F})$ is the density of states (DOS) at the Fermi level, $E_\text{F}$. As pointed out in Ref.~\citenum{Juan_2023}, the variation in the electronic structure due to charge doping is potentially one of the factors driving the magnetic instability by satisfying this condition.

Hence, the evolution of the FM order with Ni doping from the Stoner-enhanced paramagnetism of \SrCoP{} can be discussed in the framework of the Stoner model. Energy band calculations for \SrCoP{} show evidence of a sharp peak in the DOS in the vicinity of $E_\text{F}$ due to Co-3d electrons \cite{Shuyan_2025, Teruya_2014, Gotze_2021}. Within a rigid-band approximation, the development of ferromagnetic ordering with Ni doping (which adds electrons) can be understood as a consequence of the Fermi-level shift towards this DOS peak. This shift would enhance $N(E_\text{F})$, thereby satisfying the Stoner criterion. Additionally, Sr(Cu$_{1-x}$Co$_x$)$_2$P$_2$ has been investigated experimentally and discussed within the framework of the effect of electron doping, as detailed in Appendix~\ref{CuComp}.

\begin{figure}[]
 \centering
 \includegraphics[width=0.9\columnwidth]{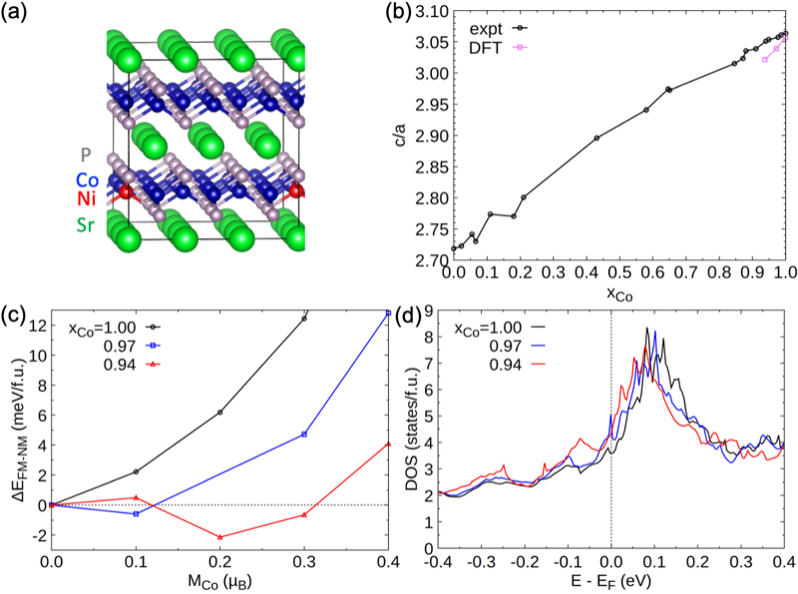}
 \caption{{Calculated magnetic properties of Sr(Ni$_{1-x}$Co$_x$)$_2$P$_2$ using the supercell approach within density functional theory (DFT). (a) The $3\times3\times1$ supercell of the conventional unit cell with one Co atom substituted for Ni, corresponding to $x_{\mathrm{Co}} = 0.97$. The $2\times2\times1$ supercell is used similarly for $x_{\mathrm{Co}} = 0.94$.
(b) The $c/a$ ratio as a function of $x_{\mathrm{Co}}$ from experiment (Ref.~\cite{Juan_2023}) and DFT calculations using the pressure-shift protocol at 12~GPa. (c) Magnetic phase stability, quantified by the DFT total energy difference between ferromagnetic (FM, fixed moment) and nonmagnetic (NM) solutions, for three compositions with increasing Ni alloying (decreasing $x_{\mathrm{Co}}$).
(d) Density of states (DOS) near the Fermi energy ($E_{\mathrm{F}}$) for the three compositions.}}
 \label{DFT}
\end{figure}

Indeed our DFT calculations, performed to understand the effect of small percentage of Ni-doping on the magnetic phase stability of \SrCoP{}, supports the development of the ferromagnetic ordering with increasing Ni concentration, as shown in Fig.~\ref{DFT}(c). For pristine SrCo$_2$P$_2$ ($x_{\mathrm{Co}}=1.00$), the FM state is unstable with $\Delta E_{\mathrm{FM-NM}}>0$ for all $M_{\mathrm{Co}}$. In contrast, slight Ni-alloying ($x_{\mathrm{Co}}=0.97$) stabilizes FM with $M_{\mathrm{Co}}=0.1\,\mu_B$ and $\Delta E_{\mathrm{FM-NM}}=-0.59$ meV/f.u, agreeing with the experimental observation. However, increasing Ni content ($x_{\mathrm{Co}}=0.94$) further enhances FM stability, yielding $M_{\mathrm{Co}}=0.2\,\mu_B$ and $\Delta E_{\mathrm{FM-NM}}=-2.14$ meV/f.u. This prediction is in clear disagreement with the experimentally observed emergence of helical AFM order. The calculations were done using the same concept as of pressure-shift protocol used in \cite{Shuyan_2025}; where the pristine supercell of \SrCoP{} is fully relaxed in the NM state giving the $c/a$ that matches the experiment, as shown in Fig.~\ref{DFT}(b). 

To further understand the electronic structure with small Ni-alloying, we plot the total NM DOS for the three compositions in Fig.~\ref{DFT}(d). With increasing Ni-alloying, the peak around $E_\mathrm{F}+0.1$~eV is shifted toward lower energy and gets gradually partially filled. Noticeably, the DOS at $E_\mathrm{F}$ also increases, which favors magnetic order in Stoner instability and qualitatively explains the observed development of the FM order upon a small amount of Ni-alloying.

Given that the DFT calculations predict an enhancement of FM stability with increasing Ni doping, its abrupt destabilization upon only a further $2$--$3\%$ Ni substitution and the concomitant stabilization of helical AFM order suggest that additional factors beyond a simple Stoner picture based solely on an enhanced density of states must be considered. An obvious explanation within the itinerant picture would be the nesting driven helical order. However, this model is yet to be verified and unlikely as it has been tested both experimentally and theoretically in related \SrNiCoAs{} compounds, and no evidence of nesting corresponding to the helical structure was found \cite{Xie_2020_SrNiCoAs}. Furthermore, other theoretical models, such as quantum order-by-disorder mechanism\cite{Xie_2020_SrNiCoAs} and frustrated one-dimensional Heisenberg model\cite{Li_2019}, previously considered for 122 systems, are incompatible here. The helical state in \SrNiCoAs{} was proposed to originate from a quantum order-by-disorder mechanism driven by itinerant-electron-mediated Ruderman-Kittel-Kasuya-Yosida (RKKY) interactions . This mechanism requires proximity to a ferromagnetic quantum-critical point (QCP), where fluctuation-induced finite-$\mathbf{q}$ instabilities pre-empt uniform ferromagnetism\cite{Thomson_2013_helicalnearQCP,Kaluarachchi_2017}. In contrast, the helical phase in our system emerges on the opposite side of the ferromagnetic "dome", away from QPT (or putative QCP), indicating that a different microscopic origin is at play, and also suggests that the mechanism in Ni-doped SrCo$_2$As$_2$ may warrant further scrutiny. Additionally, the frustrated one-dimensional, as discussed below in Appendix~\ref{ODHM}, provides only a qualitative description at best as the increase of the turn angle, as shown in Fig.~\ref{AFM_t-x_dep} (b), with increasing $x$ (on approaching FM from the helical side) seems inconsistent with the model. The observed increase in the turn angle is more consistent with an evolution toward A-type magnetic order, whereas the stabilization of an FM state would require a dramatic reversal of this trend and an abrupt collapse of the turn angle to $0^\circ$. This leaves open the key question of which subtle band-structure or exchange parameters evolve with doping to suppress ferromagnetism and stabilize the helical state.

Overall, several conventional frameworks, including Stoner enhancement, order-by-disorder mechanisms, and the frustrated 1D Heisenberg model, used in related 122 systems capture certain qualitative features of the phase diagram but fail to account for the full evolution of the magnetic ground states observed here. These discrepancies underscore that the itinerant magnetism in this system cannot be described by these previously established paradigms and remains a theoretical question. 

\section{Conclusion}

In summary, we have investigated the evolution of magnetic ordering in \SrNiCoP{} series through single-crystal neutron diffraction measurements on compositions spanning the AFM to FM regimes. For \(x = 0.88\) and \(0.94\), our results establish an incommensurate helical ground state with a doping-dependent propagation vector \((0,0,\tau)\), closely resembling the behavior previously reported in \SrNiCoAs{}. In contrast, the \(x = 0.97\) composition exhibits clear signatures of long-range ferromagnetic order. Together with previous magnetization results, our measurements unambiguously establish the presence of bulk FM order in this composition. Consistent with these earlier results, our results also show a monotonic decrease of ordered the moment as ordering temperature decreases and $x$ approaches $1.0$.

The emergence and evolution of these distinct magnetic ground states highlight the delicate balance of competing interactions in this system. Notably, the observed behavior is not readily captured by conventional itinerant or local-moment Heisenberg models that have been applied to related 122 materials, including frameworks based on Stoner model, order-by-disorder, or quasi-one-dimensional Heisenberg descriptions. This discrepancy underscores the need for a more refined theoretical treatment that incorporates the effects of subtle band-structure changes accompanying Ni–Co substitution. 

\begin{acknowledgments}
	Work at Ames National Laboratory was supported by the U.\,S.\ Department of Energy, Office of Basic Energy Sciences, Division of Materials Sciences \& Engineering. Ames National Laboratory is operated for the U.\,S.\ Department of Energy by Iowa State under Contract No.\ DE-AC$02$-$07$CH$11358$. A portion of this research used resources at the High Flux Isotope Reactor, a U.\,S.\ Department of Energy, Office of Science User Facility operated by Oak Ridge National Laboratory. The The beam time was allocated to VERITAS and PTAX instruments on proposal numbers IPTS-25125.1 and IPTS-27873.1, respectively. Some of the computation used resources of the National Energy Research Scientific Computing Center (NERSC), a DOE Office of Science User Facility.
\end{acknowledgments}

\section {Appendix:}
\subsection{Single Crystal Growth and Characterization\label{SCGC}}
Single crystals of Sr(Ni$_{1-x}$Co$_x$)$_2$P$_2$ and of Sr(Cu$_{1-x}$Co$_x$)$_2$P$_2$ were obtained by high-temperature solution growth method \cite{Canfield_2020}
out of Sn flux. High-purity elements were loaded into a $2\ \text{ml}$ alumina fritted Canfield Crucible Set \cite{Canfield_2016_fritCrucible,LSPCeramics}, and sealed under partial atmosphere of Argon in a fused silica tube. 

For growing Sr(Ni$_{1-x}$Co$_x$)$_2$P$_2$, we followed the procedure given at Ref. \cite{Juan_2023}, with a starting ratio Sr:(Ni+Co):P:Sn of 1.2:2:2:20 for the samples with low Ni fraction explored in this work. The ampules were placed inside a box
furnace, held for 6 hours at $600\ ^{\circ}\text{C}$ before increasing to $1180\ ^{\circ}\text{C}$, dwelled for 24 hours to make sure the material was fully melted, and finally slowly cooled down over 100 hours to $1000\ ^{\circ}\text{C}$, the temperature at which the excess high-temperature solution was decanted with the aid of a centrifuge. 

For the case of Sr(Cu$_{1-x}$Co$_x$)$_2$P$_2$, we could only grow samples with compositions very close to either parent compound (SrCu$_2$P$_2$ or SrCo$_2$P$_2$), but not the intermediate compositions. Initially, we prepared several growths combining the elements in a ratio Sr:(Cu+Co):P:Sn of 1.2:2:2:20, using different proportions of Co and Cu and the same temperature sequence described above. Upon increasing the Cu:Co ratio in the melt, we obtain crystals of Sr(Cu$_{1-x}$Co$_x$)$_2$P$_2$ crystals with an decreasing $x$ until reaching $x\leq0.989(2)$ for a critical Cu:Co ratio of 0.7:0.3, beyond which the target phase no longer grows and instead we obtain a mixture of other phases (CoSn$_3$, Co$_2$Sn and SrCuP) at lower temperatures (900$^{\circ}$C). We proceeded, therefore, to lower the amounts of Sr and Sn and increase the amounts of P in the initial melt, with a ratio of 1.1:2:2.15:18, being able to obtain crystals with $x\leq0.979(3)$. Still, at Cu:Co ratios beyond 0.7:0.3, the other phases grew instead as well. When decreasing the Sr/Sn and increasing the P even further, to a ratio 
of 1:2:2.3:16, and extending the cooling down to 800$^{\circ}$C, we were able to obtain Sr(Cu$_{1-x}$Co$_x$)$_2$P$_2$ crystals with $x\leq 0.012$. We were not able to grow crystals with an intermediate range of compositions, $0.012\leq x\leq0.979$, despite trying other initial ratios of the elemental compounds. Additionally, we attempted the self-flux method analogous to the one reported for Sr(Ni$_{1-x}$Rh$_x$)$_2$P$_2$, with a starting ratio Sr:(Cu+Co):P of 5:48:47, which resulted in the growth of CoP only. Increasing the ratio of Sr led to an increase of the melting point of the mixture beyond the softening point of the silica ampules. 

Finally, we attempted to obtain intermediate compositions in polycrystalline form by solid-state reaction, combining Sr:(Cu+Co):P in the ratio 1:2:2 in alumina crucibles placed inside sealed silica ampules under a partial atmosphere of argon, which was heated to 250$^{\circ}$C, held for 6 hours, then to 600$^{\circ}$C, held for another 6 hours, then to 800$^{\circ}$C, held for 6 hours, and finally to 1180$^{\circ}$C and held for 24 hours. The ampule was then allowed to cool outside the furnace. The solidified mixture was ground and pressed into pellets, and placed in an alumina crucible inside a sealed ampule, which was kept at 1180$^{\circ}$C for 100 hours. After this, the mixture was allowed to cool and solidify outside the furnace, and the mixture was ground and pressed into a pellet again. Part of the ground powder was used to measure powder x-ray diffraction, in order to follow the evolution of the solid-state reaction. This procedure was repeated three times. No single phase was obtained in any of the attempts made, and at most we identified a mixture of both Cu-rich and Co-rich Sr(Cu$_{1-x}$Co$_x$)$_2$P$_2$ phases, together with other binary compounds (primarily CoP). This result suggests that there may be an immiscibility gap inherent to the system that prevents the intermediate compositions with $0.012\leq x\leq0.979$ to form. 

The atomic fractions of each element in the crystals were determined by Energy Dispersive x-ray Spectroscopy (EDS) quantitative chemical analysis using an EDS detector (Thermo NORAN Microanalysis System, model C10001) attached to a JEOL scanning-electron microscope (SEM). An acceleration voltage of
$22\ \text{kV}$, working distance of $10\ \text{mm}$ and take off angle of $35 ^{\circ}$ were used for measuring all standards and crystals with unknown composition. A single crystal of SrCo$_2$P$_2$ was used as standard for Sr, Co and P quantification, a SrNi$_2$P$_2$ crystal as a standard for Ni, and a SrCu$_2$P$_2$ crystal as a standard for Cu. The Cu fractions were also compared to those obtained when using GdCu$_2$ as a standard for Cu, obtaining consistent results. The spectra were fitted using NIST-DTSA II Microscopium software \cite{Newbury2014}. The composition of each platelike crystal was measured at different positions on the crystal's face (perpendicular to $c$), as well as different points across a polished edge of the crystal (along $c$). The average compositions and error bars were obtained from these data, accounting for both inhomogeneity and goodness of fit of each spectra. 

Powder x-ray diffraction (XRD) measurements were performed using a Rigaku MiniFlex II powder diffractometer with Cu K$\alpha$ radiation ($\lambda=1.5406\ \text{\AA}$). For each composition, a few crystals were finely ground to powder and dispersed evenly on a single crystal Si zero background holder, with the aid of a small quantity of vacuum grease. Intensities were collected for $2\theta$ ranging from $10^{\circ}$ to $100^{\circ}$, in step sizes of $0.01^{\circ}$, counting for 4 seconds at each angle. Rietveld refinement was performed using GSAS II software package \cite{Toby_2013_GSAS}. Refined parameters included but were not limited to phase fractions, lattice parameters, atomic positions and isotropic displacements.

\subsection{Comparison with Sr(Cu$_{1-x}$Co$_x$)$_2$P$_2$\label{CuComp}}

\begin{figure}[]
 \centering
 \includegraphics[width=\columnwidth]{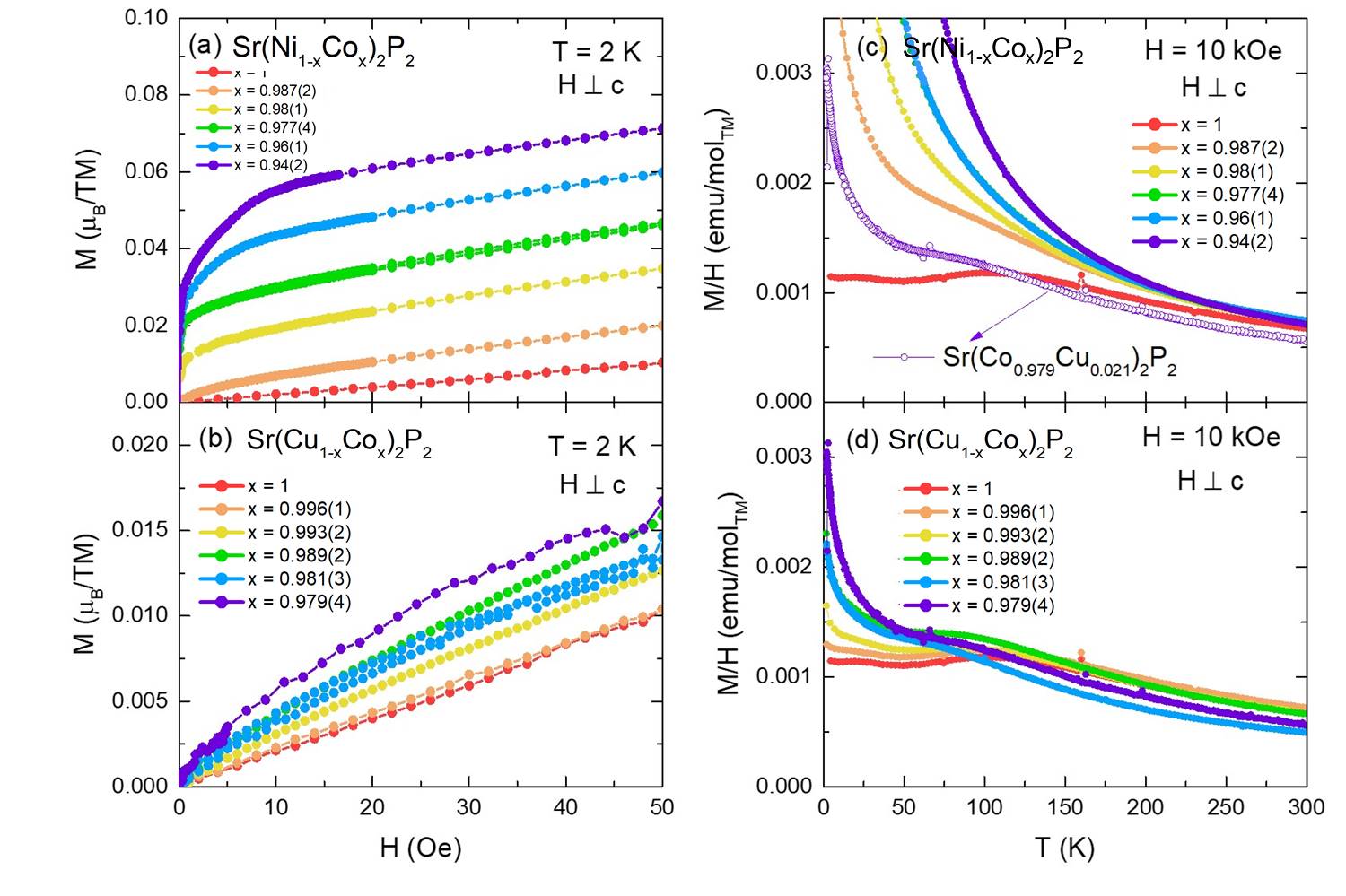}
 \caption{{(a)–(b) Field-dependent magnetization for Sr(Ni$_{1-x}$Co$_x$)$_2$P$_2$ and Sr(Cu$_{1-x}$Co$_x$)$_2$P$_2$ with $H \perp c$, demonstrating that the FM component present in \SrCoNiP{} is absent in the Cu-doped samples. (c)–(d) Temperature-dependent magnetization normalized by an applied field of 10 kOe ($H \perp c$) for the same series. The purple symbols in (c) correspond to the highest Cu-substituted sample obtained in our growth, included for direct comparison with the Ni-doped samples; despite having the largest magnetization among Sr(Cu$_{1-x}$Co$_x$)$_2$P$_2$, it remains lower than all Sr(Ni$_{1-x}$Co$_x$)$_2$P$_2$ compositions.}}
 \label{Mag_SrCoTMP}
\end{figure}

The success of the DFT calculations in capturing the evolution of ferromagnetic (FM) order in \SrNiCoP{}, an interpretation based solely on rigid-band shifts for Stoner instability, together with changes in the electronic structure induced by variations in the lattice parameters ($c/a$), might suggest naively that it should also apply to Cu doping: Cu substitution for Co would likewise be expected to shift the bands with increasing DOS at $E_\mathrm{F}$, and, experimentally, produces a similar trend in $c/a$, which should then favor a ferromagnetic instability. Furthermore, since Cu ($3d^{10}4s^1$) nominally contributes more electrons per substituent than Ni ($3d^{8}4s^2$), one might even anticipate a more pronounced band-shift effect in the Cu-doped case. However, the experimental results shown in Fig.~\ref{Mag_SrCoTMP} demonstrate that Cu doping does not stabilize a ferromagnetic ground state in this system. As seen in Figs.~\ref{Mag_SrCoTMP}(a) and (b), the spontaneous magnetization observed in the Ni-doped series for $x \geq 0.02$ is entirely absent in the Cu-doped samples studied. Even though samples with higher Cu substitution could not be grown, the sample with highest Cu fraction ($x=0.979$) would have a similar electron count as the Co-substituted sample with $x=0.96$ that shows clear FM behavior. Furthermore, the temperature-dependent magnetization normalized per $TM$ atom [Figs.~\ref{Mag_SrCoTMP}(c) and \ref{Mag_SrCoTMP}(d)] shows that Cu substitution enhances the low-temperature magnetization by nearly an order of magnitude less than Ni substitution at comparable doping levels.

Hence, the lack of observed magnetic order in Sr(Cu$_{1-x}$Co$_x$)$_2$P$_2$ for $0\leq x \leq 0.979$ may indicate that the electron count by itself is not a sufficient explanation for the emergent magnetism in Sr(Ni$_{1-x}$Co$_x$)$_2$P$_2$. However, this comparison should be treated with caution, as Cu behaves less like a transition metal and instead acts more like an alkali-group substitution. Also, as discussed in Ref.~\cite{Wadati_2010}, it should not be used for simple electron count.

\begin{figure}[h]
 \centering
 \includegraphics[width=\columnwidth]{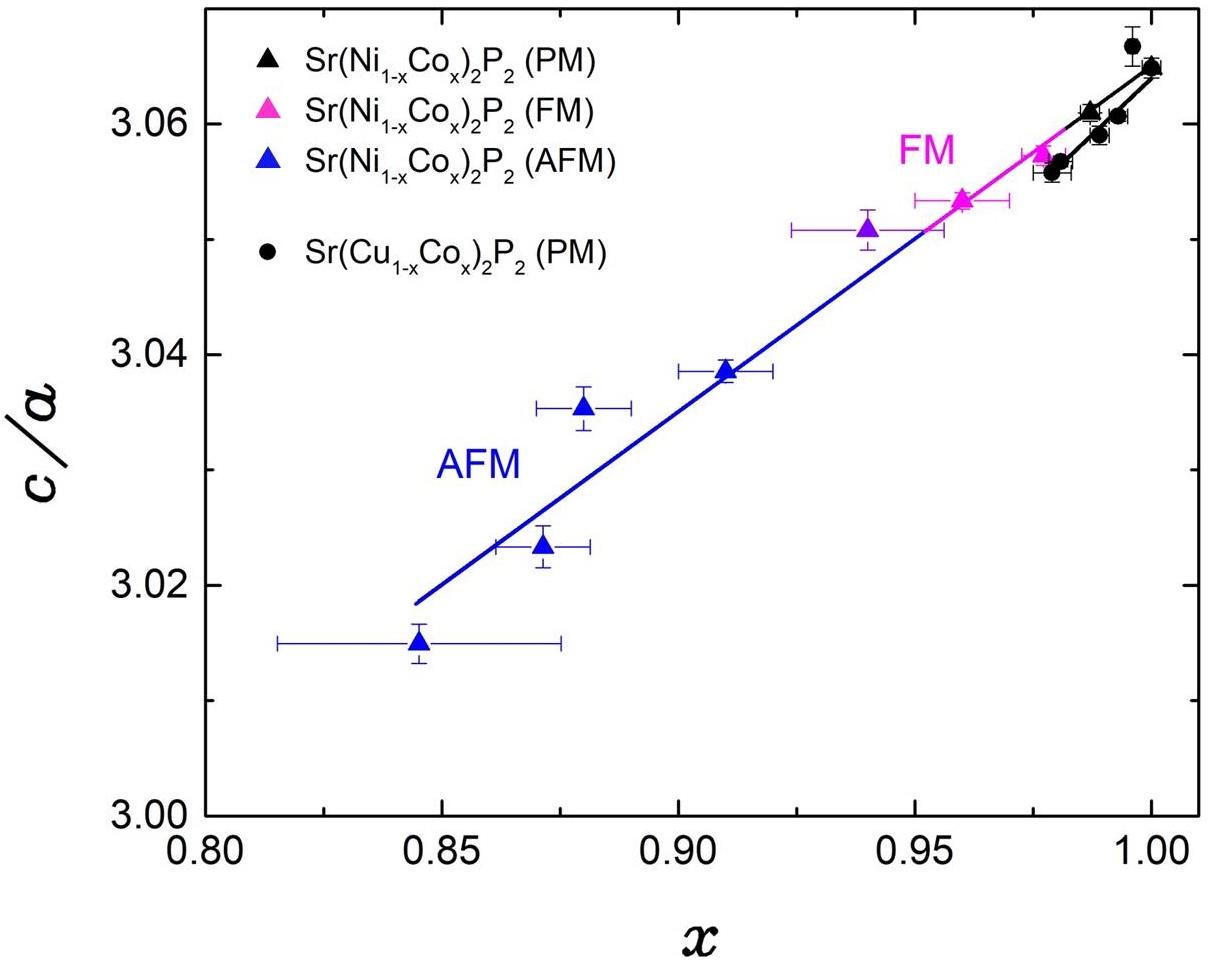}
 \caption{{Evolution of $c/a$ ratio as a function of composition, $x$, in Sr(Ni$_{1-x}$Co$_x$)$_2$P$_2$ (triangles) and Sr(Cu$_{1-x}$Co$_x$)$_2$P$_2$ (circles). The pink and turquoise triangles correspond to the ferromagnetic and antiferromagnetic Sr(Ni$_{1-x}$Co$_x$)$_2$P$_2$ samples, respectively. The black symbols correspond to samples that exhibit paramagnetism down to 1.8~K.}}
 \label{fig:c_a_SrTM2P2}
\end{figure}

Some reports have claimed that the relevant tuning parameter leading to emergent magnetism could be the $c/a$ ratio of the crystal structure. In order to address this, Fig. \ref{fig:c_a_SrTM2P2} compares the evolution of $c/a$ for Sr(Ni$_{1-x}$Co$_x$)$_2$P$_2$ (triangles) and Sr(Cu$_{1-x}$Co$_x$)$_2$P$_2$ (circles). Both display similar trends and Sr(Cu$_{1-x}$Co$_x$)$_2$P$_2$ reaches low enough $c/a$ values comparable with FM compositions of Sr(Ni$_{1-x}$Co$_x$)$_2$P$_2$. In spite of all this, no magnetic order is observed in Sr(Cu$_{1-x}$Co$_x$)$_2$P$_2$, suggesting that $c/a$ cannot be the leading tuning parameter for magnetism in these systems. 

\subsection{One-dimensional Heisenberg model\label{ODHM}}
In the one-dimensional Heisenberg model, competition between the interlayer nearest-neighbor (NN) exchange $J_{z}$ and next-nearest-neighbor (NNN) exchange $J'_{z}>0$ (AFM) yields FM, helical, and A-type magnetic ground states. In the absence of anisotropy, the helical phase occupies $-4 < J_{z}/J'_{z} < 4$, with turn angle
\[
\phi = \cos^{-1}\!\left[-\frac{J_{z}}{4J'_{z}}\right].
\]
Anisotropy further suppresses the helical region in favor of collinear states. Although this frustrated-chain picture captures the qualitative competition between magnetic states in \SrNiCoP{}, it fails to reproduce the experimental evolution of the turn angle. Specifically, with increasing $x$, i.e., on approaching the FM state, the turn angle within the helical phase increases toward $180^\circ$ (Fig.~\ref{AFM_t-x_dep}), whereas the model predicts a continuous decrease toward $0^\circ$ as the FM boundary is approached. This behavior instead resembles an approach toward A-type order, such that the abrupt transition from a large-angle helical state to FM is unexpected within a simple $J_{z}$--$J'_{z}$ framework. Thus, while the frustrated-chain model offers qualitative guidance, it does not capture the observed evolution of the turn angle in this itinerant system.

\bibliographystyle{apsrev4-2.bst}
\bibliography{SrNiCoP.bib}


\end{document}